\documentclass[english]{emulateapj}

\def\gs{\mathrel{\raise1.16pt\hbox{$>$}\kern-7.0pt %  
\lower3.06pt\hbox{{$\scriptstyle \sim$}}}}         %  
\def\ls{\mathrel{\raise1.16pt\hbox{$<$}\kern-7.0pt %  
\lower3.06pt\hbox{{$\scriptstyle \sim$}}}}         %   

\slugcomment{ApJ, in press}
\shorttitle{The Origin of the 24$\rm \mu m$ Excess in Red Galaxies}
\shortauthors{Brand et al.}

\makeatother
\begin{document}

\title{The Origin of the 24$\rm \mu m$ Excess in Red Galaxies}

\author{Kate Brand\altaffilmark{1}, John Moustakas\altaffilmark{2},
  Lee Armus\altaffilmark{3}, Roberto ~J. Assef\altaffilmark{4},
  Michael ~J.~I. Brown\altaffilmark{5}, Richard
  ~R. Cool\altaffilmark{6}, Vandana Desai\altaffilmark{3}, Arjun
  Dey\altaffilmark{6}, Emeric Le Floc'h\altaffilmark{7}, Buell
  ~T. Jannuzi\altaffilmark{6}, Christopher
  ~S. Kochanek\altaffilmark{4}, Jason Melbourne\altaffilmark{8}, Casey
  ~J. Papovich\altaffilmark{9}, B.~T. Soifer\altaffilmark{4,8}}

\altaffiltext{1}{Space Telescope Science Institute, 3700 San Martin Drive, Baltimore, MD 21218; brand@stsci.edu}
\altaffiltext{2}{Center for Cosmology and Particle Physics, New York University, 4 Washington Place, New York, NY 10003}
\altaffiltext{3}{Spitzer Science Center, California Institute of Technology, 220-6, Pasadena, CA 91125}
\altaffiltext{4}{Department of Astronomy, Ohio State University, Columbus, OH43210}
\altaffiltext{5}{School of Physics, Monash University, Clayton, Victoria 3800, Australia}
\altaffiltext{6}{Steward Observatory, Tucson, AZ 85721}
\altaffiltext{6}{National Optical Astronomy Observatory, 950 North Cherry Avenue, Tucson, AZ 85726} 
\altaffiltext{7}{Institute for Astronomy, University of Hawaii, 2680 Woodlawn Drive, Honolulu, HI 96822, USA}
\altaffiltext{8}{Division of Physics, Mathematics and Astronomy, California Institute of Technology, 320-47, Pasadena, CA 91125}
\altaffiltext{9}{Department of Physics, Texas A\&M University, 4242 TAMU, College Station, TX 77843}

\begin{abstract}
Observations with the $Spitzer~Space~Telescope$ have revealed a
population of red sequence galaxies with a significant excess in their
24$\rm \mu m$ emission compared to what is expected from an old
stellar population. We identify $\sim$900 red galaxies with $0.15 \le
z \le 0.3$ from the AGN and Galaxy Evolution Survey (AGES) selected
from the NOAO Deep Wide-Field Survey Bo\"otes field. Using $Spitzer$
MIPS, we classify 89 ($\sim$10\%) with 24$\rm \mu m$ infrared excess
(f$_{24} \ge$0.3mJy). We determine the prevalence of AGN and
star-formation activity in all the AGES galaxies using optical line
diagnostics and mid-IR color-color criteria. Using the IRAC
color-color diagram from the IRAC Shallow Survey, we find that 64\% of
the 24$\rm \mu m$ excess red galaxies are likely to have strong PAH
emission features in the 8$\rm \mu m$ IRAC band. This fraction is
significantly larger than the 5\% of red galaxies with f$_{24}
<$0.3mJy that are estimated to have strong PAH emission, suggesting
that the infrared emission is largely due to star-formation
processes. Only 15\% of the 24$\rm \mu m$ excess red galaxies have
optical line diagnostics characteristic of star-formation (64\% are
classified as AGN and 21\% are unclassifiable). The difference between
the optical and infrared results suggest that both AGN and
star-formation activity is occurring simultaneously in many of the
24$\rm \mu m$ excess red galaxies. These results should serve as a
warning to studies that exclusively use optical line diagnostics to
determine the dominant emission mechanism in the infrared and other
bands. We find that $\sim$40\% of the 24$\rm \mu m$ excess red
galaxies are edge-on spiral galaxies with high optical
extinctions. The remaining sources are likely to be red galaxies whose
24$\rm \mu m$ emission comes from a combination of obscured AGN and
star-formation activity.
\end{abstract}

\keywords{galaxies: elliptical and lenticular, cD --- galaxies: starburst --- infrared: galaxies --- quasars: general}

\section{Introduction}
\label{sec:intro}

The optical color-magnitude relation of galaxies shows a clear bimodality, with a narrow sequence of red (predominantly spheroidal) galaxies and a ``blue cloud'' of actively star-forming galaxies (e.g., \citealt{str01}, \citealt{hog04}). Studies of the evolution of the red galaxy luminosity function suggest that the stellar mass within the red sequence has increased since $z\sim 1$, perhaps by a factor of $\approx2-4$ (\citealt{bel04}; \citealt{fab07}; \citealt{bro07}). The most likely cause of this growth is the transition of blue cloud galaxies to red sequence galaxies at $z<1$. Recent semi-analytic models of galaxy formation support such a scenario (e.g., \citealt{cro06}). In order to reproduce the color bimodality, one or more mechanisms must be introduced to shut down the star-formation (e.g., \citealt{som99}), and turn blue galaxies red. Some authors suggest that feedback from merger-induced AGN activity could be the quenching mechanism (e.g., \citealt{hop06}, \citealt{geo08}), while others suggest that supernovae in a starburst galaxy may have the required energy to blow out enough gas \citep{ben03}. Other possible mechanisms include virial shock heating of the galaxy halo \citep{bir07} or simple gas consumption. If we are to understand how galaxies shut off their star-formation and become red, it is clearly important to identify and study the population of sources that have recently moved onto the red sequence or are in a transition region between the red sequence and blue cloud commonly referred to as the ``green valley''. These sources may still show signs of AGN and/or residual star-formation as they move onto the red sequence. 
 
Optical studies of the stellar populations of red galaxies generally show them to be dominated by an old component (e.g., \citealt{tin68}), with perhaps a ``frosting'' of stars formed since $z$=1 (e.g., \citealt{tra05}). In studies of local ellipticals, the weak infrared emission is usually attributed to mass loss from the evolved stellar population (\citealt{tem05}; \citealt{bre06}). However, there has been recent evidence from $Chandra$ and $Spitzer$ that both hidden AGN and star-formation activity may occur in red galaxies out to $z\approx 1$ (e.g., \citealt{bra05}, \citealt{rod07}, \citealt{dav06}). More than half of all red galaxies have narrow emission lines in their optical spectra indicative of AGN activity (e.g., \citealt{yan06}; \citealt{gra07}).  

We have identified a population of red galaxies with relatively strong 24$\rm \mu m$ emission (f$_{24}>$0.3 mJy; hereafter ``24$\rm \mu m$ excess'' red galaxies). The redshift range 0.15$\le z\le$0.3 allows us to use both optical and infrared diagnostics to determine whether they are likely to be dominated by AGN or star-formation activity. We use starburst templates \citep{dal02} to estimate the total infrared luminosities of L$_{8-1000\rm \mu m}\ge 1 \times 10^{10} ( 3 \times 10^{10}) \rm~ L_\odot$ at $z$ = 0.15 (0.3). Assuming that all the dust heating is due exclusively to star formation processes and adopting the \citet{sal55} IMF (0.1-100 $M_\odot$), this corresponds to SFRs of $\sim$2 (5) M$_\odot \rm ~yr^{-1}$ \citep{ken98}. This infrared luminosity is orders of magnitude larger than expected from local ellipticals (which have infrared luminosities from 5$\times 10^{7}$ - 7$\times 10^{8} \rm~ L_\odot$; \citealt{dev99}) and suggests optically obscured activity in these galaxies. Similar results are found by \citet{rod07} for $0.3<z<1$ spheroidal galaxies in GOODS \citep{dic03} and by \citet{dav06} for z$\le 0.3$ red galaxies in SWIRE \citep{lon03}. 

In this paper, we investigate the nature of the 24$\rm \mu m$ excess in these galaxies. We use mid-IR colors to determine whether the sources have significant polycyclic aromatic hydrocarbon (PAH) emission features that are known to be strong in luminous star-forming galaxies. We also use optical line diagnostics to distinguish between photo-ionization by young massive stars in star-forming HII regions and by AGN. The nature of the 24$\rm \mu m$ emission from optically selected AGN and other populations will be explored further in W. Rujopakarn in preparation.

A cosmology of $H_0 = 70 {\rm ~km ~s^{-1} ~Mpc^{-1}}$, $\Omega_M$=0.3, and $\Omega_\Lambda$=0.7 is assumed throughout. All optical magnitudes are on the Vega system. 

\section{Data}

Our sample is selected from the AGN and Galaxy Evolution Survey (AGES; C.~S. Kochanek et al. in prep.; J. Moustakas et al. in prep.). AGES is an optical spectroscopic survey conducted within the Bo\"otes field of the NOAO Deep Wide-Field survey (NDWFS; \citealt{jan99}). The survey was carried out in 2004 and 2005 using the Hectospec multi-fiber spectrograph (\citealt{fab98}; \citealt{fab05}) at the MMT 6.5m telescope. Successful redshifts were measured for nearly 20,000 objects. For our analysis, we use the AGES ``main galaxy sample'', which is statistically complete for $15<I<19.95$, after correcting for sparse sampling, fiber incompleteness, and redshift failures (see Eisenstein et al., in preparation for details). The AGES sparse sampling preferentially targets brighter optical galaxies and/or sources detected in other wavelengths and results in roughly half of the total $I<$19.95 sample with good quality optical spectra. We account for any selection effects in which one type of source is selected more frequently than another by applying a correcting weight to each galaxy (each individual galaxy is weighted by the inverse of the known sampling rate of galaxies with similar properties). The final percentages of red, green, blue and 24 $\rm \mu m$ excess red galaxies with different optical and infrared classifications only changes by a few percent at the most.

The optical spectroscopy covers the wavelength range 3700-9200\AA\ at a FWHM resolution of $\sim$6\AA. The AGES spectra were reduced using standard procedures implemented in ${\small HSRED}$, a customized Hectospec reduction package. The rest-frame optical emission lines were modeled using Gaussian line profiles, after carefully subtracting the underlying stellar continuum using the \citet{bru03} population synthesis models; additional details regarding the AGES data reduction and emission line measurements can be found in J. Moustakas et al. in preparation. 

The Bo\"otes field covers a contiguous 9.3 square degree area and has a plethora of existing multi-wavelength imaging and spectroscopy. The field has been mapped in $B_W$, $R$, and $I$ bands to median 3$\sigma$ point source depths of $\approx$27.7, 26.7, and 26.0 (Vega) respectively. The {\em Spitzer~Space~Telescope} \citep{wer04} has imaged the field at 24, 70, and 160 $\rm \mu m$ using the Multiband Imaging Photometer for $Spitzer$ (MIPS; \citealt{rie04}) to 5$\sigma$ rms depths of 0.3 mJy, 25 mJy, and 150 mJy respectively, yielding a catalog of $\approx$22,000 sources. $Spitzer$ has also imaged the field with the Infrared Array Camera (IRAC; \citealt{faz04}) to 5$\sigma$ depths of 6.4, 8.8, 51, and 50 $\mu$Jy at 3.6, 4.5, 5.8, and 8$\rm \mu m$ respectively \citep{eis04}.

\section{Source Selection}

We have selected a sample of 3889 galaxies from the AGES survey that have spectroscopic redshifts, $0.15\le z < 0.3$ and absolute V-band magnitudes of M$_{\rm v} \le $-19.4 mag. This redshift range allows the IRAC color-color diagram to be used as a diagnostic for the presence of PAH emission (with the strong PAH emission features redshifted into the observed 8$\rm \mu m$ band) and ensures that we cover the rest-frame wavelengths of several powerful emission line diagnostics, including  H$\alpha$, H$\beta$, [OIII] $\lambda\lambda$4959,5007, and [NII] $\lambda\lambda$6548,6584. At absolute magnitudes fainter that M$_{\rm v} >$-19.4 mag, the number of red galaxies decreases significantly. Our absolute optical magnitude cut ensures that we are comparing red and blue galaxies within a similar absolute magnitude range. The optical magnitude cut also ensures that these are galaxies selected from the main AGES sample (with $I<$19.95) that has easily quantifiable selection effects. The rest-frame optical luminosities and colors were computed using ${\small K-CORRECT}$ (v. 4.1.4; \citealt{bla03}; \citealt{bla07}). For more details, the reader is referred to J. Moustakas et al. in preparation.

We classify the galaxies into ``red sequence'' (914 sources) and ``blue cloud'' (2255 sources) using the following rest-frame selection criteria:\\

Red Sequence:
\begin{equation}
\begin{array}{ll}
% U - V > &  ~1.55 - 0.25 - 0.04 \times (M_V + 20.0) \\
%        &  - 0.42 \times (z - 0.05) + 0.07 \times (z - 0.05)^2  
 U - V > &  ~1.3 - 0.04 \times (M_V + 20.0) - 0.42 \times (z - 0.05). 
\end{array}
\end{equation}

Blue Cloud:
\begin{equation}
\begin{array}{ll}
 U - V \le &  ~1.3 - 0.04 \times (M_V + 20.0) - 0.42 \times (z - 0.05).  
\end{array}
\end{equation}

\begin{figure}[h]
\begin{center}
\includegraphics[height=80mm]{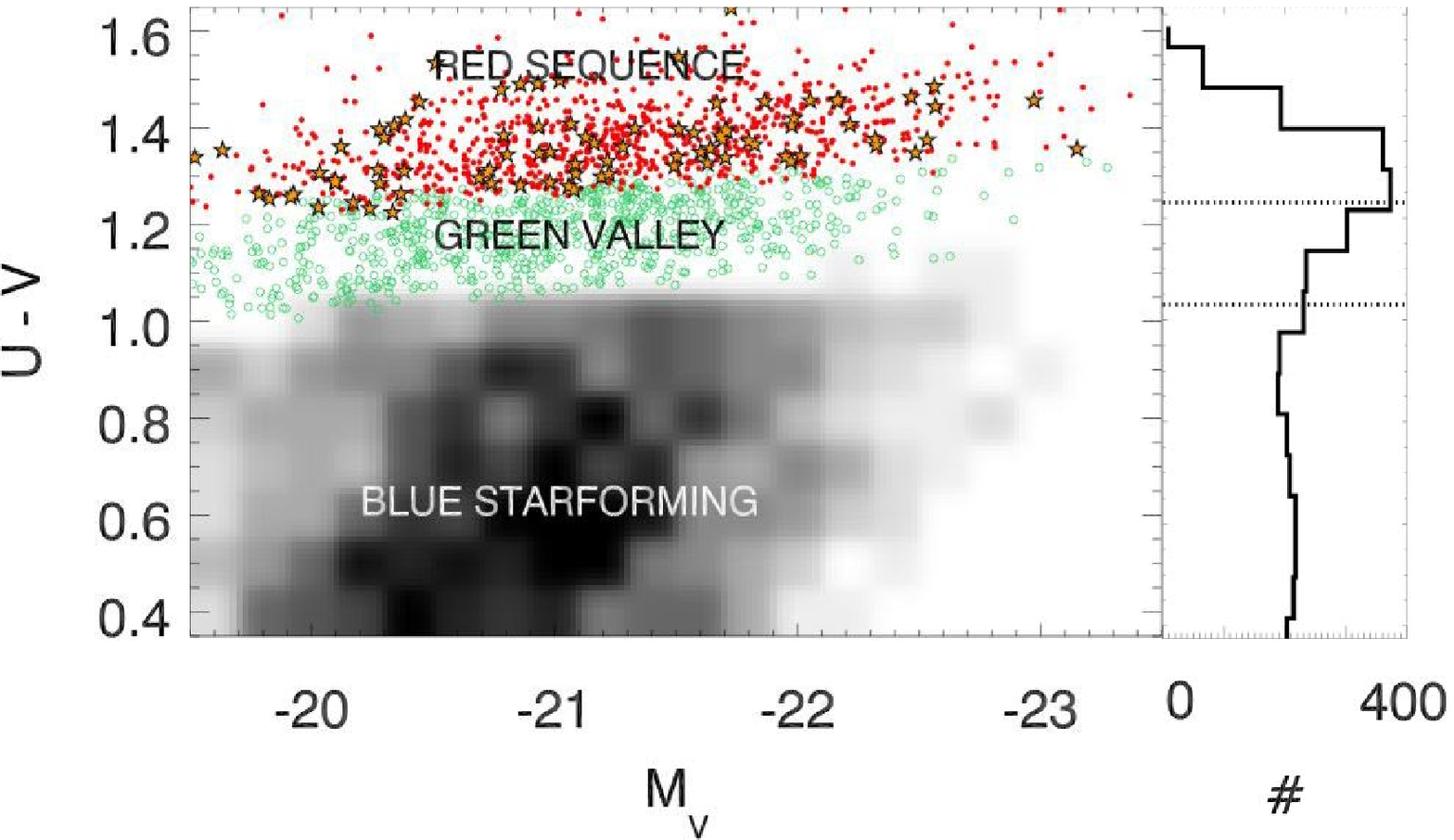}
\end{center}
\caption{\label{fig:colmag1} Color-magnitude diagram for all $M_{V}<$-19.4 mag AGES galaxies at $0.15<z<0.3$. The sources are divided into red sequence galaxies (filled red circles), green valley galaxies (empty green circles), blue cloud galaxies (greyscale). Also shown is a histogram of the distribution of U-V colors of the sources. The dotted lines show the selection criteria at $M_{V}$=-20 mag and $z$=0.225.}
\end{figure}

The 720 sources that fall between these criteria on the color-magnitude diagram are classified as ``green valley'' sources. Figure~\ref{fig:colmag1} shows the color-magnitude diagram for the entire sample divided into the different classifications. The selection criterion for red sequence galaxies is similar to that of \citet{bro07}, although we adopt a slightly redder cut to limit contamination and do not employ any optical or infrared apparent magnitude color cuts. 

Of the 914 red galaxies, 89 ($\sim$10\%) have 24 $\rm \mu m$ flux
densities, f$_{24}>$0.3mJy ([24$\rm \mu m$]$<$16.1 mag). We refer to
this population as ``24$\rm \mu m$ excess'' red galaxies (red galaxies
with f$_{24}<$0.3 mJy are referred to as ``24$\rm \mu m$ faint'' red
galaxies). All of these have IRAC 3.6 $\rm \mu m$ detections and we
expect all 24 $\rm \mu m$ matches to be that of a true counterpart (we
expect only 0.07\% of our galaxies to have 24 $\rm \mu m$ matches
within a radius of 2\arcsec of the 24$\rm \mu m$ position by
coincidence). We note that our use of f$_{24}>$0.3mJy to define
``24$\rm \mu m$ excess'' red galaxies is a somewhat arbitrary limit
determined by the depth of our MIPS 24$\rm \mu m$ data. It does
however correspond to infrared luminosities of L$_{8-1000 \rm \mu
  m}\ge 1 \times 10^{10} ( 3 \times 10^{10}) \rm~ L_\odot$ at $z$ =
0.15 (0.3) that are orders of magnitude higher than we would expect
from an old stellar population (see Section~\ref{sec:intro}).

\section{Results}

In this Section, we use infrared colors, optical line diagnostics, optical morphology, and mean multi-wavelength spectral energy distributions to investigate the nature of the 24$\rm \mu m$ excess red galaxies. Only 8/914 ($<$1\%) of the red galaxies have X-ray detections in the XBo\"otes Chandra survey (\citealt{ken05}; \citealt{bra06}). The X-ray survey limit corresponds to an X-ray luminosity of 5$\times 10^{41}$ (2 $\times 10^{42}$) ergs s$^{-1}$ at $z$=0.15 (0.3). The small number of detections suggests that any X-ray emission from AGN is generally either weak or obscured and so we do not discuss the X-ray emission further. 

\subsection{The IRAC Color-Color Diagram}
\label{sec:irac}

\begin{figure}[h]
\begin{center}
\includegraphics[height=80mm]{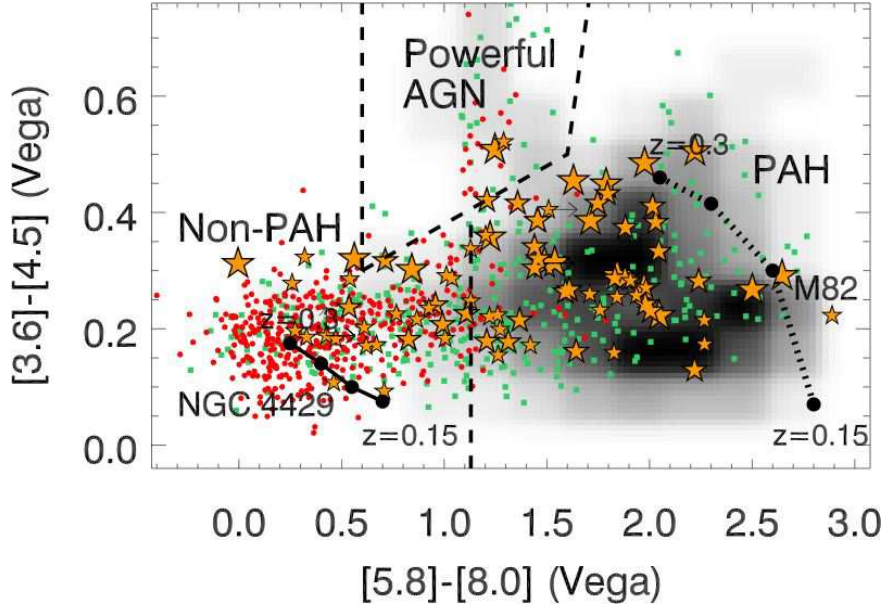}
\end{center}
\caption{\label{fig:irac} IRAC color-color diagram for all AGES galaxies.  The greyscale shows the distribution of blue galaxies. The red circles, green squares, and yellow stars denote red, green valley, and 24$\rm \mu m$ excess red galaxies respectively. The larger yellow stars correspond to 24$\rm \mu m$ excess red galaxies with larger 24$\rm \mu m$ flux densities. For clarity, for the blue, green, and red galaxies, only the sources with detections in all 4 IRAC bands are shown. For the 24$\rm \mu m$ excess red galaxies, where an IRAC flux is below the detection limit, the limits on the IRAC colors are shown. Over-plotted are the non-evolving 0.15$\le z <$0.3 color tracks for M82 (a starburst galaxy; dotted black line) and NGC 4429 (a massive inactive S0/Sa galaxy; solid black line). In both cases, the redshift increases upwards and to the left. The diagram is divided into three sections by the dashed black line: the powerful AGN ``wedge'' as defined by \citet{ste05}, and the ``PAH'' and ``Non-PAH'' regions defined empirically from the distribution of the red and blue galaxies.}
\end{figure}

The IRAC color-color diagram (e.g., \citealt{lac04}; \citealt{ste05};
\citealt{gor08}) has been shown to be a reliable way of identifying
high redshift AGN. Figure~\ref{fig:irac} shows the IRAC color-color
diagram for the AGES galaxies at 0.15$\le z <$0.3. The vast majority
of sources have [3.6]-[4.5] colors consistent with them being low
redshift galaxies (e.g., \citealt{ste05}). The red and blue galaxies
show a striking bimodality in [5.8]-[8.0] color (this is also
demonstrated in \citealt{ass08}). The blue galaxies have redder
[5.8]-[8.0] colors which is likely due to the strong 6.2 and 7.7 $\rm
\mu m$ PAH emission features falling within the observed 8 $\rm \mu m$
band-pass at these redshifts (e.g., \citealt{smi07}). The red galaxies
are likely to have bluer [5.8]-[8.0] colors because of the weakness or
lack of PAH emission features and/or the 1.6$\rm \mu m$ stellar
photospheric feature (the ``stellar bump'') that may still dominate
the infrared emission in less active sources with large old stellar
populations. We overplot the color evolution of M82 and NGC 4429 using
the templates of \citet{dev99}. M82 is a starburst galaxy with strong
PAH emission features. NGC 4429 is an inactive massive S0/Sa galaxy in
the Virgo cluster. The tracks confirm that actively star-forming and
non-star-forming galaxies lie in different regions of the IRAC
color-color diagram at these redshifts.

\setcounter{footnote}{0}

\begin{figure}[h]
\begin{center}
\includegraphics[height=80mm]{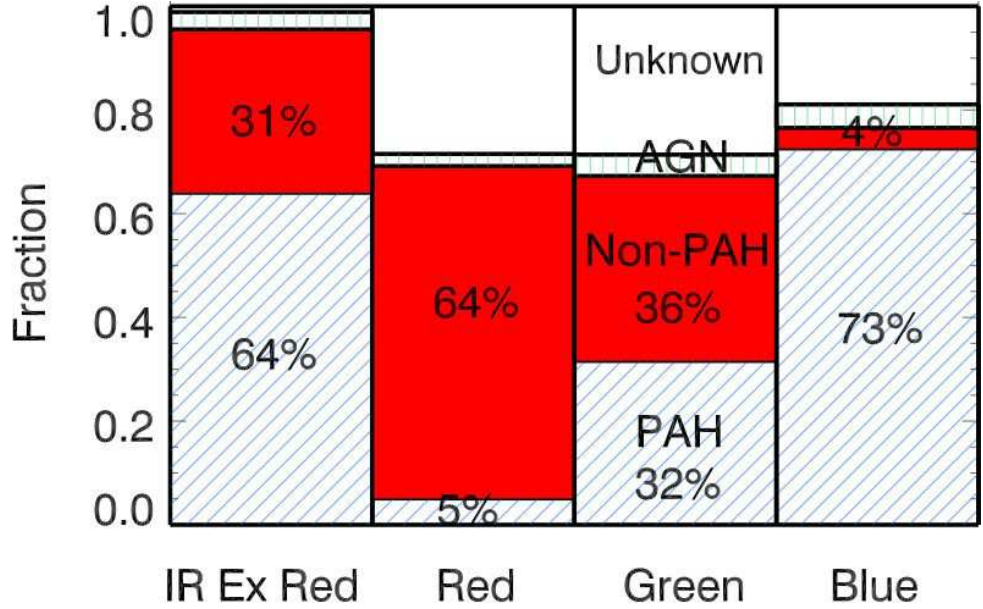}
\end{center}
\caption{\label{fig:hist} Classification of galaxies based on the IRAC color-color diagram (Figure~\ref{fig:irac}). The histogram shows the fraction of 24$\rm \mu m$ excess red galaxies, 24$\rm \mu m$ faint red galaxies, green galaxies, and blue galaxies within the PAH (blue slanted lines), Non-PAH (solid red shading), and powerful AGN (green vertical lines) regions or that are unclassifiable (``Unknown''; white area). The fractions have been corrected for the AGES incompleteness. The fractions of 24$\rm \mu m$ excess red galaxies, 24$\rm \mu m$ faint red galaxies, green galaxies, and blue galaxies within the AGN (unclassifiable) regions are 3\% (2\%), 2\% (29\%), 4\% (28\%), and 5\% (18\%) respectively.
}
\end{figure}

The 24$\rm \mu m$ excess red galaxies lie in the region of the
color-color diagram between that of the blue star-forming galaxies and
the general red galaxy population (and in a similar region to that of
the green valley sources). There is no correlation of [5.8]-[8.0]
color with f$_{24}$. The infrared brightest sources fall in both the
non-PAH and PAH regions. To quantify the distribution of sources in
Figure~\ref{fig:irac}, we divide the color-color space into 3 regions:
the ``AGN'' wedge region expected to be populated by powerful AGN as
defined by \citet{ste05}, and the ``PAH'' and ``Non-PAH'' regions. The
boundary between these two populations at an infrared color of
[5.8]-[8.0] = 1.13 mag is defined empirically so that only $\sim$5\%
of the blue galaxies are defined as non-PAH sources and $\sim$5\% of
the non-24$\rm \mu m$ excess red galaxies are defined as PAH
sources. To be classified, we require sources to have a detection in
at least one IRAC band and to appear unambiguously in one region of
the IRAC color-color diagram once limits are taken into account
(otherwise, it is classified as ``unknown''). Figure~\ref{fig:hist}
shows the fraction of red, green, blue, and infrared-excess red
galaxies classified as powerful AGN, PAH, or non-PAH from
figure~\ref{fig:irac}. Approximately 64\% of the 24$\rm \mu m$ excess
red galaxies are classified as PAH sources (compared to only 5\% of
the 24$\rm \mu m$ faint red galaxies\footnote{We note that this number
  is likely to be slightly higher as 28\% of the 24$\rm \mu m$ faint
  red galaxies are unclassifiable. However, this cannot account for
  the large difference, even in the unlikely event that all of the
  remaining 28\% were classified as PAH sources.}). This result
suggests that the infrared emission is largely due to star formation.

\subsection{The Optical Line Diagnostic Diagram}

\begin{figure}[h]
\begin{center}
\includegraphics[height=80mm]{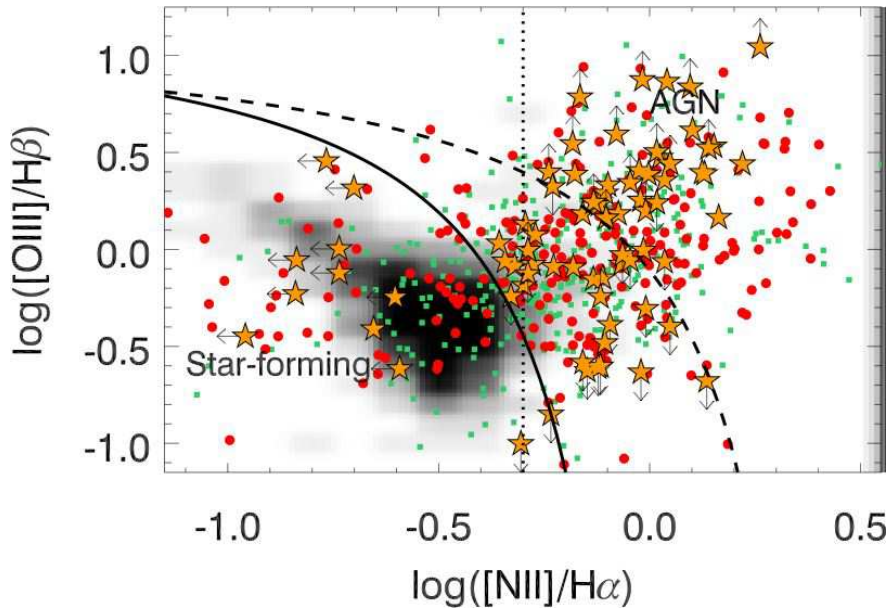}
\end{center}
\caption{\label{fig:bpt} The log([OIII]$\lambda$5007/H$\beta$) vs. log([NII]$\lambda$6584/H$\alpha$) line diagnostic diagram for galaxies with measurable line fluxes. The symbols are as defined in Figure~\ref{fig:irac}. The empirically derived line of \citet{kau03} (solid line) used to distinguish between star-forming galaxies and narrow-line AGN is shown as a solid line. The \citet{kew01} extreme starburst classification line (dashed line) is also shown for reference. The line at log([NII]/H$\alpha$=-0.3 used to classify sources with no suitable H$\beta$ and/or [NII] measurement is also shown (dotted line). For the 24$\rm \mu m$ excess galaxies, arrows show the limits on the line ratios in cases where lines are not detected.}
\end{figure}

 In Figure~\ref{fig:bpt} we show an optical line diagnostic diagram (\citealt{bal81}; \citealt{kew01}; \citealt{kau03}; \citealt{kew06}) for all of the AGES galaxies. We consider all 1996 sources with H$\alpha$ and [NII]$\lambda$6584 emission lines with a signal to noise ratio of S/N$>$3 (51\% of the total sample). If a source has an optical spectrum with S/N$>$3 for all four of the H$\alpha$, H$\beta$, [OIII]$\lambda$5007, and [NII]$\lambda$6584 emission lines, it is classified as an AGN or star-forming galaxy using the empirically derived classification line of \citet{kau03}. We also include sources with a signal to noise of S/N$>$3 on the H$\alpha$ and [NII] emission lines but with S/N$<$3 on either of the H$\beta$ and [OIII] emission lines. We classify these sources as an AGN if log([NII]/H$\alpha >$ -0.3 and a star-forming galaxy if log([NII]/H$\alpha \le$ -0.3 (see \citealt{tre04} for a similar classification criterion). We classify sources with S/N$>$3 on the H$\alpha$ and S/N$<$3 on the [NII] emission lines and log([NII]$_{lim}$]/H$\alpha$) $\le$ -0.3 (where [NII]$_{lim}$ is the 1$\sigma$ upper limit on the [NII] emission line measurement) as star-forming galaxies. Any source with S/N$>3$ on both the H$\alpha$ and H$\beta$ emission lines and FWHM$>$500 km $s^{-1}$ is classified as a (broad-line) AGN. If a source does not meet any of these criteria,  it is classified as ``unknown''. This results in a completeness rate of 70/89 (79\%), 241/825 (29\%), 377/720 (52\%), and 1942/2255 (86\%) for 24$\rm \mu m$ excess red galaxies, 24$\rm \mu m$ faint red galaxies, green galaxies, and blue galaxies respectively at 0.15$\le z <$0.3. 

Blue galaxies tend to have [OIII]/H$\beta$ versus [NII]/H$\alpha$ line ratios expected for star-forming galaxies whereas red galaxies lie in a much larger region, spanning all classifications. The 24$\rm \mu m$ excess red galaxies fall in a region similar to that of the green valley galaxies. 

\begin{figure}[h]
\begin{center}
\includegraphics[height=80mm]{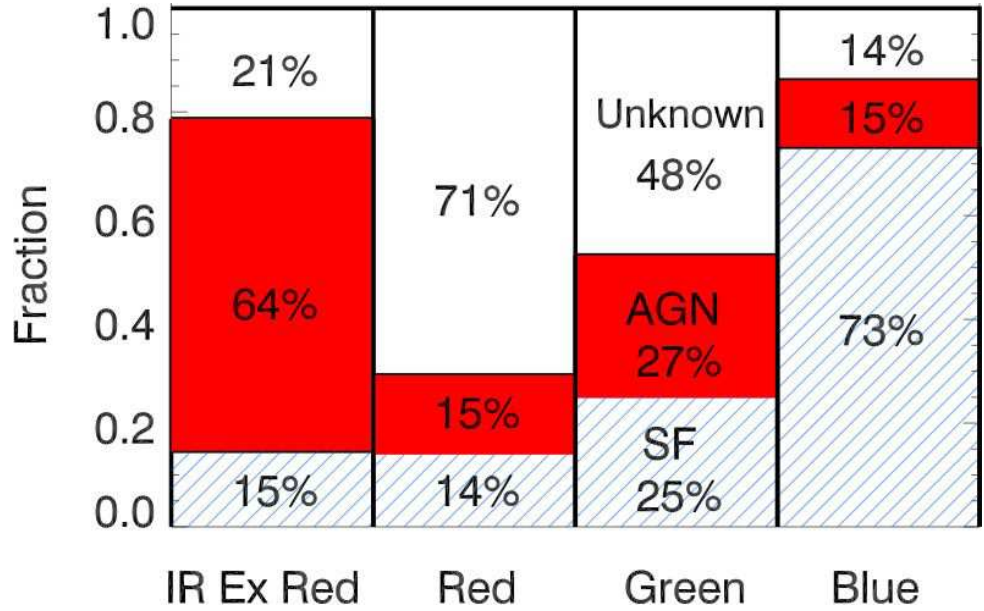}
\end{center}
\caption{\label{fig:bpt_hist}  Classification of galaxies based on their optical line diagnostics (see Figure~\ref{fig:bpt}). The histogram shows the fraction of 24$\rm \mu m$ excess red galaxies, 24$\rm \mu m$ faint red galaxies, green galaxies, and blue galaxies within the star-forming (``SF'';blue slanted lines) and AGN (``AGN'';solid red shading) regions of the log([OIII]/H$\beta$) vs. log([NII]/H$\alpha$ line diagnostic diagram  and sources whose optical spectra are not good enough to classify the sources (``unknown''; white area). The fractions have been corrected for the AGES incompleteness.}
\end{figure}

Figure~\ref{fig:bpt_hist} shows the fraction of 24$\rm \mu m$ excess red galaxies, 24$\rm \mu m$ faint red galaxies, green galaxies, and blue galaxies classified as star-forming, AGN, or unknown (and with the sparse sampling weights taken into account). Galaxies with broad optical lines are included in the AGN classification. 57/89 (64\%) and 13/89 (15\%) of the 24$\rm \mu m$ excess red galaxies are classified as AGN and star-forming sources respectively (with 19/89 (21\%) unclassifiable).

The 24$\rm \mu m$ excess red galaxies with infrared colors consistent with PAH emission features are not confined to the star-forming region but are distributed throughout the optical line diagnostic diagram. This result suggests that a large proportion of 24$\rm \mu m$ excess red galaxies exhibit characteristics of both AGN and star-forming galaxies. 

\subsection{The Multi-Wavelength SED}

\begin{figure}[h]
\begin{center}
\includegraphics[height=90mm]{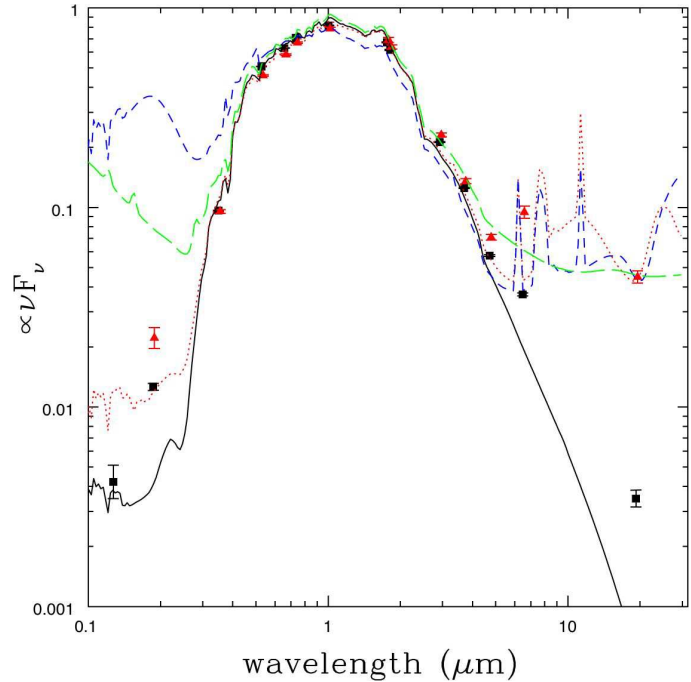}
\end{center}
\caption{\label{fig:sed} The average multi-wavelength spectral energy distribution of the 24$\rm \mu m$ excess red galaxies (red triangles) and 24$\rm \mu m$ faint red galaxies (black squares). Also shown are the errors on the mean flux density. Overplotted are SED templates from \citet{ass08}. The solid black line represents an elliptical galaxy template normalized to the peak of the emission from the 24$\rm \mu m$ faint red galaxies. The dotted red line (dashed blue line; long-dashed green line) shows the sum of the Sbc spiral template (Irr template; AGN template) and the elliptical galaxy template that simultaneously match the 24$\rm \mu m$ emission of the 24$\rm \mu m$ excess red galaxies and the peak of the stellar emission.} 
\end{figure}

Figure~\ref{fig:sed} shows the mean multi-wavelength spectral energy distribution (SED) of the 24$\rm \mu m$ excess red galaxies compared to that of the 24$\rm \mu m$ faint red galaxies. Overplotted are the SED templates of \citet{ass08} for elliptical, spiral, irregular galaxies and AGN extended into the UV and infrared (these modifications will be presented by Assef et al. in preparation). The far and near-UV points are obtained from a Galaxy Evolution Explorer (GALEX; \citealt{mar05}) survey in the NDWFS Bo\"otes field. The mean SEDs are essentially indistinguishable except in the infrared (by definition) and the UV. The excess of UV emission in the 24$\rm \mu m$ excess red galaxies is likely due to the same processes that are giving rise to the 24$\rm \mu m$ excess. The mean SED of the 24$\rm \mu m$ faint red galaxies is well fit by an elliptical galaxy template (the slight excess in the near-UV and at 24$\rm \mu m$ can be accounted for by the fact that this class still includes some sources with a modest 24$\rm \mu m$ excess which do not satisfy the requirement of f$_{24}\ge$0.3mJy that they would need to be classified as a 24$\rm \mu m$ excess red galaxy). The mean UV to 24$\rm \mu m$ flux density ratio of the 24$\rm \mu m$ excess red galaxies is a factor of 10(4) smaller than that of the irregular galaxy (AGN) template and twice as high as that of the Sbc spiral galaxy template. This suggests that they contain a larger amount of dust and/or host an older stellar population than typical irregular galaxies. If the 24 $\rm \mu m$ and UV excesses are due largely to AGN, then the UV emission must be suppressed by dust. The amount of dust is not well constrained: an extinction of E(B-V) of a few tenths would be sufficient to suppress the UV light.

\subsection{The Color Magnitude Diagram}

\begin{figure}[h]
\begin{center}
\includegraphics[height=80mm]{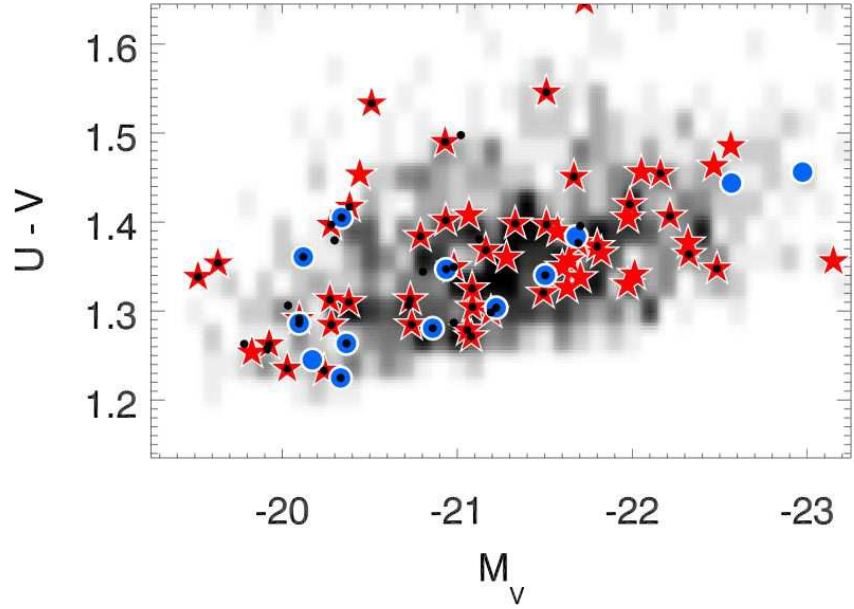}
\end{center}
\caption{\label{fig:colmag} Color-magnitude diagram showing rest-frame $U$-$V$ color vs absolute $V$-band magnitude (Vega). The greyscale shows the distribution of all red sequence galaxies. All symbols represent 24$\rm \mu m$ excess red galaxies. Filled red stars and filled blue circles denote 24$\rm \mu m$ excess red galaxies classified as AGN and star-forming galaxies respectively from the optical line diagnostic diagram (\citealt{bal81}; \citealt{kau03}; \citealt{kew06}; see Figure~\ref{fig:bpt}). The black dots denote 24$\rm \mu m$ excess red galaxies classified as having PAH emission from the IRAC color-color diagram (see Figure~\ref{fig:irac}).
}
\end{figure}

Figure~\ref{fig:colmag} shows the distribution of the 24$\rm \mu m$ excess red galaxies in color-magnitude space compared to that of the general red galaxy population. A Kolmogorov-Smirnov test shows that the distribution of the 24$\rm \mu m$ excess red galaxies (i.e., all colored points in Figure~\ref{fig:colmag}) is not significantly different from that of the overall red galaxy population. However, there are hints of differences in the distribution of 24$\rm \mu m$ excess red galaxies with different classifications from the optical line diagnostic diagram: the star-forming generally have bluer $U-V$ colors than the AGN sources. 

\subsection{Optical Morphology}
\label{sec:morph}

We use the NDWFS $R$-band images to investigate the optical morphologies of the 24$\rm \mu m$ excess red galaxies. We find that for 42\% of the 24$\rm \mu m$ excess red galaxies, the optical images exhibit an elongated structure (the remaining sources tend to resemble typical bulge-dominated galaxies). The large axial ratios ($\sim$ 5-10:1) are indicative of an edge-on spiral galaxy (hereafter referred to as a ``spiral morphology''). The fraction of 24$\rm \mu m$ excess red galaxies with a spiral morphology is much larger than the 11\% of 24$\rm \mu m$ faint red galaxies with such morphologies and suggests that some fraction of the 24$\rm \mu m$ excess red galaxies are reddened edge-on disk galaxies. Of the 24$\rm \mu m$ excess red galaxies within the PAH region of the infrared color-color diagram, 50\% have spiral morphologies (compared to 32\% of the ``non-PAH'' sources).

We also compare the local environments of the 24$\rm \mu m$ excess red galaxies to that of the 24$\rm \mu m$ faint red galaxies. The 24$\rm \mu m$ excess galaxies show no evidence for a larger fraction of sources with either a disturbed morphology or close companion compared to that of the 24$\rm \mu m$ faint red galaxies. 

\section{Discussion}

We find that a small but significant minority (10\%) of red galaxies
have an 24$\rm \mu m$ excess. We have used the IRAC colors of 24$\rm
\mu m$ excess red galaxies to show that a large fraction are likely to
exhibit PAH emission features, indicative of star-formation. Although
the infrared emission of elliptical galaxies is usually attributed to
AGN or stellar mass loss from evolved stars (e.g., \citealt{tem05};
\citealt{bre06}), PAH emission features have recently been detected in
nearby elliptical galaxies \citep{kan05}. \citet{kan05} find that the
6.2, 7.7, and 8.6 $\rm \mu m$ emission features are unusually weak in
comparison to PAH emission features at longer
wavelengths. \citet{smi07} show that this peculiar PAH emission
spectrum with markedly diminished 5-8 $\rm \mu m$ features arises
among the sample solely in systems with relatively hard radiation
fields harboring low-luminosity AGN. Given this effect, it is perhaps
even more surprising that we find such a large fraction of 24$\rm \mu
m$ excess red galaxies in the ``PAH'' region of the IRAC color-color
diagram.

There appears to be a discrepancy between the optical and infrared
diagnostics of the 24$\rm \mu m$ excess red galaxies: the optical line
diagnostics suggest that AGN dominate whereas the infrared colors
suggest that star formation dominates. How do we reconcile these
differences? This could be, at least in part, an aperture effect. The
optical spectra are obtained using a multi-fiber spectrograph and the
fibers subtend 1.5\arcsec on the sky. This corresponds to an angular
diameter of $\approx$5.4 kpc at $z$=0.225. The IRAC magnitudes were
obtained from 6\arcsec diameter apertures, corresponding to angular
diameter of $\approx$22 kpc at $z$=0.225. The optical diagnostics are
therefore likely to probe only the region of the galaxy bulge, whereas
the infrared diagnostics consider infrared light from the entire
galaxy. Probing only the central regions may result in a larger
contribution by an AGN, if present. The optical line diagnostics are
also very sensitive to AGN activity. In particular, \citet{kew00} have
shown that 80\% of optically classified AGN have compact radio cores,
demonstrating that optical spectroscopy identifies AGN with a high
rate of success, despite potential problems with dust
obscuration. Presumably, in many cases, the extended narrow-line
emission is detected even when the nucleus is heavily dust obscured
and the infrared colors are dominated by PAH emission from
circum-nuclear star formation. It is also likely that different
processes may dominate at different wavelengths. This study provides a
caution against using diagnostics in one wavelength to determine what
dominates the bolometric luminosity of the source.

What is the nature of the 24$\rm \mu m$ excess red galaxies? We explore three possibilities:\\

(1) Dusty star-forming galaxies with high extinctions. The 24$\rm \mu m$ excess red galaxies could be a population of dusty star-forming galaxies with enough extinction to give them redder $U$-$V$ colors (e.g., \citealt{ala02}; \citealt{unt08}). In Section~\ref{sec:morph}, we show that 42\% of the 24$\rm \mu m$ excess red galaxies have optical morphologies indicative of edge-on spiral galaxies.This fraction is much larger than the 11\% of 24$\rm \mu m$ faint red galaxies with such morphologies (and the $\sim$15\% of the red sequence population estimated to be dusty galaxies in the HST GEMS survey at $z\sim$0.7 \citep{bel04}). We compute the H$\alpha$/H$\beta$ Balmer decrements for all 24$\rm \mu m$ excess red galaxies with a signal to noise, S/N$>$3 on the H$\alpha$ and H$\beta$ lines (48/89=54\% of the total sample). Of the measurable sources, we find that $\approx$60\% have optical extinctions that are sufficiently large that their intrinsic optical colors are consistent with them being blue or green valley galaxies (compared to only 13\% of the 24$\rm \mu m$ faint red galaxies with measurable lines). Similar results were reported by \citet{dav06}, who find that $\sim$50\% of their infrared bright red galaxies classified as star-forming galaxies have intrinsic optical colors consistent with them being blue sequence galaxies.  However, the large bulge components that they find in their sources suggest that these are not pure disc-like galaxies with high levels of extinction. Of the 24$\rm \mu m$ excess red galaxies within the PAH region of the infrared color-color diagram, 50\% have spiral morphologies (compared to 32\% of the ``non-PAH'' sources). This result suggests that dusty star-forming galaxies contribute to some, but not all, of the 24$\rm \mu m$ excess red galaxy population.

(2) Quenched galaxies that have recently joined the red sequence from the blue cloud. The 24$\rm \mu m$ excess red galaxies could represent a transition population of sources whose star-formation has been recently quenched (in $<$1 Gyr; \citealt{hop06}; \citealt{sch07}). The population exhibits characteristics that would be consistent with such a scenario. One could envisage a situation in which the 24$\rm \mu m$ excess red galaxies that show star-forming characteristics in their optical line diagnostics are the sources that are in the earlier stages of being quenched. As the sources become quenched further, they become redder in their $U$-$V$ colors, and their optical line diagnostics become dominated by AGN.  In Figure~\ref{fig:colmag}, we show the color magnitude diagram for the 24$\rm \mu m$ excess red galaxies split into star-forming sources and AGN as classified from their optical line diagnostics. The star-forming 24$\rm \mu m$ excess red galaxies appear to be slightly bluer on average than the sources classified as AGN. \\

(3) Merging red galaxies. The 24$\rm \mu m$ excess red galaxy population could be massive red galaxies that are undergoing major or minor mergers and exhibiting obscured star-formation and AGN activity. While the optical line diagnostics suggest that low level AGN activity appears to be common in both 24$\rm \mu m$ excess and 24$\rm \mu m$ faint red galaxies, the 24$\rm \mu m$ excess red galaxies may indicate a recent galaxy merger or interaction that triggers obscured star-formation activity that is only detectable in the infrared (and marginally in the UV which is more sensitive to star-formation activity than the optical). This may lower the need for ``dry'' mergers in massive red galaxies. The distribution of stellar masses (estimated from the optical photometry) for the 24$\rm \mu m$ excess red galaxies ($\approx 10^{10}-10^{11.5} \rm M_\odot$) shows no significant difference from that of the 24$\rm \mu m$ faint red galaxies. In addition, the infrared luminosity is poorly correlated with the stellar mass estimated from the optical photometry (J. Moustakas et al., in preparation), consistent with what we would expect if there are brief episodes of activity occurring within the same parent population. These results are more consistent with the 24$\rm \mu m$ excess population being dominated by merging red galaxies rather than quenched galaxies that have recently joined the red sequence. In the latter case, the ``downsizing'' of galaxies (in which galaxies of decreasing mass are progressively joining the red sequence with time) would imply that the masses of 24$\rm \mu m$ excess red galaxies should be less than that of 24$\rm \mu m$ faint red galaxies (that joined the red sequence at a higher redshift). On the other hand, we show in Section~\ref{sec:morph} that the 24$\rm \mu m$ excess galaxies show no evidence for a larger fraction of sources with either a disturbed morphology or close companion than that of the 24$\rm \mu m$ faint red galaxies.\\

It appears that the low redshift 24$\rm \mu m$ excess red galaxy population comprises a mixture of different types of sources. Roughly 40\% are likely to be edge-on disk galaxies with high extinction. The rest are likely to be red galaxies whose 24$\rm \mu m$ emission comes from both obscured AGN and star formation activity. They may be a transition population of sources that have only recently left the blue cloud to join the red sequence and in which the residual star-formation is being quenched by AGN activity that becomes increasingly dominant in the optical band as the sources become redder. Alternatively, they could be massive red galaxies undergoing a burst of obscured star-formation activity (perhaps due to a merger). Further work needs to be done to determine which of these mechanisms dominates and how this changes with redshift.

\section{Conclusions}

By analyzing the infrared colors and optical line diagnostics of 0.15$\le z \le$0.3 24$\rm \mu m$ excess red galaxies and comparing these to that of blue, green valley, and 24$\rm \mu m$ faint red galaxies within the AGES survey, we find the following:

\begin{itemize}
\item{Of the total red galaxy population, 10\% have 24 $\rm\mu m$ flux densities, f$_{24}>$0.3 mJy. These correspond to infrared luminosities that are larger than we would expect from the infrared emission of local elliptical galaxies.}
\item{Using their [5.8]-[8.0] IRAC colors, we show that 64\% of the
  24$\rm \mu m$ excess red galaxies are likely to have strong PAH
  emission features in their infrared spectra (compared to only 5\% of
  the 24$\rm \mu m$ faint red galaxies). This result suggests that in
  a large fraction of the sources, the infrared emission is dominated
  by star-formation processes.}
\item{A much larger fraction of 24$\rm \mu m$ excess red galaxies are classified as star-forming using infrared diagnostics ($\sim$64\%) than optical line diagnostics ($\sim$15\%). This result suggests that in many sources, both AGN and star-formation activity are occurring simultaneously (but that in the optical, the AGN dominates the signal whereas in the infrared, the PAH emission features dominate.)}
\item{Although the optical to near-IR SEDs of the 24$\rm \mu m$ excess and 24$\rm \mu m$ quiet galaxies are very similar, the 24$\rm \mu m$ excess red galaxies appear to also have a small UV excess over that of the 24$\rm \mu m$ quiet galaxies. The small UV to 24$\rm \mu m$ ratio suggests a large amount of dust and/or relatively old stellar population compared to that of irregular galaxies.}
\item{Approximately 40\% of 24$\rm \mu m$ excess red galaxies at low redshifts are likely to be dusty edge-on spiral galaxies, whose high optical extinctions make them red. The rest are likely to be red galaxies whose 24$\rm \mu m$ emission comes from both obscured AGN and star formation activity. This may be triggered by minor mergers of red galaxies or, in some cases, may be the residual activity as blue galaxies are quenched to form red sequence galaxies.}
\end{itemize}

We thank our colleagues on the NDWFS, MIPS, AGES, and IRAC teams. KB
is supported by the Giacconi fellowship at STScI. JM acknowledges
funding support from NASA-06-GALEX06-0030 and Spitzer G05-
AR-50443. This research is partially supported by the National Optical
Astronomy Observatory which is operated by the Association of
Universities for Research in Astronomy, Inc. (AURA) under a
cooperative agreement with the National Science Foundation. This work
is based on observations made with the {\it Spitzer~Space~Telescope},
which is operated by the Jet Propulsion Laboratory, California
Institute of Technology under a contract with NASA. The {\it Spitzer /
  MIPS} survey of the Bo\"otes region was obtained using GTO time
provided by the {\it Spitzer} Infrared Spectrograph Team (James Houck,
P.I.) and by M. Rieke. AGES is a collaboration between scientists at
the Harvard-Smithsonian Center for Astrophysics, Steward Observatory,
the National Optical Astronomical Observatory, the Ohio State
University, and the Jet Propulsion Laboratory.We thank the anonymous
referee for his/her useful comments.

%\bibliography{ms}

\end{document}